\documentclass[useAMS,usenatbib]{mn2e}
\usepackage{graphicx}
\usepackage{epstopdf}
\usepackage{subfigure}
\usepackage[a4paper]{hyperref}
\usepackage{amssymb}
\usepackage{psfig}

 \def\be{\begin{equation}}
 \def\ee{\end{equation}}
\def\solmas{{{\rm M}$_\odot$}}
\def\simless{\mathbin{\lower 3pt\hbox
   {$\rlap{\raise 5pt\hbox{$\char'074$}}\mathchar"7218$}}}   % < or of order
\def\simgreat{\mathbin{\lower 3pt\hbox
   {$\rlap{\raise 5pt\hbox{$\char'076$}}\mathchar"7218$}}}   % > or of order
\def\etal{{\rm et al.}}

\def\solm{{\rm M}_\odot}

\def\tff {t_{\rm ff}}

\title[Clustered and distributed star formation] {The efficiency of star formation in clustered and distributed regions}

\author[I. A. Bonnell \etal]
  {Ian A. Bonnell$^1$\thanks{E-mail: iab1@st-and.ac.uk}, Rowan J. Smith$^{1,2}$, Paul C. Clark$^2$,  and Matthew R. Bate$^3$  \\
$^1$ SUPA, School of Physics and
  Astronomy, University of St Andrews, North Haugh, St Andrews, Fife,
  KY16 9SS. \\
$^2$  Institut fuer Theoretische Astrophysik, Albert-Ueberle-Str. 2, 
69120, Heidelberg, Germany \\
$^3$ School of Physics, University of Exeter, Stocker Road, Exeter, EX4 4QL \\ }

\date{\today}

\begin{document}

\date{Accepted 0000 December 00. Received 0000 December 00; in original form 0000 October 00}

\pagerange{\pageref{firstpage}--\pageref{lastpage}} \pubyear{2009}

\maketitle

\label{firstpage}

\begin{abstract}
We investigate the formation of both clustered and distributed populations of young stars in a single molecular cloud.
We present a numerical simulation of a $10^4 \solm$ elongated, turbulent,  molecular cloud and the formation of over 2500 stars. The
stars form both in stellar clusters and in a distributed mode which is determined by the local gravitational binding of the cloud. 
A density gradient along the major axis of the cloud produces bound regions that form stellar clusters and unbound regions that
form a more distributed population.  The initial mass function also depends on the local gravitational binding of the cloud
with bound regions forming full IMFs whereas in the unbound, distributed regions the stellar masses
cluster around the local Jeans mass and lack both the high-mass and the low-mass stars. The overall efficiency
of star formation is $\approx 15$\% in the cloud when the calculation is terminated, but varies from less than 1\% in the the regions of distributed star formation to
$\approx  40$\% in regions containing large stellar clusters.  Considering that large scale surveys are likely
to catch clouds at all evolutionary stages, estimates of the (time-averaged) star formation efficiency for the giant molecular cloud reported here is only $\approx 4$\%. This would lead to the erroneous conclusion of {\sl slow} star formation when in fact it
is occurring on a dynamical timescale.
\end{abstract}


\section{Introduction}

The ability to conduct wide-area surveys of molecular clouds  has shown
that most stars form in clusters containing some hundreds to thousands of stars (\citealt{Ladaetal1991}; \citealt{ClaBonHil2000}; \citealt{LadLad2003}). At the same time, mid-infrared surveys such
as Spitzer have shown that significant numbers of stars form in a more distributed mode \citep{AllenMegeathetal2007, Gutermuthetal2008, Gutermuthetal2009, Evansetal2009}.
The reason why such different modes of star formation exist, and in the same cloud (e.g.\ Orion A) is unclear.

There has also been considerable interest as to why star formation appears to be inefficient \citep{Evansetal2009},
with only a few percent of a molecular cloud's mass being turned into stars per free-fall time. This could imply
that star formation is a slow process \citep{KruTan2007} or that it is an inherently inefficient process, but
proceeds on the local dynamical timescale. In the
latter case the efficiency must increase on small scales where bound clusters are formed. For example, the Orion
Nebula Cluster has a median age of $\approx 10^6$ years and a dynamical time of $\approx 3 \times 10^5$ years
\citep{HilHar1998}.
Given an overall star formation efficiency of $\approx 50$\% this implies a star formation per free-fall time of 15\%.
Considering that the initial pre-cluster cloud is likely to have been at least a factor of 2 larger \citep{BonBatVin2003}, this
implies an efficiency of star formation per {\sl initial} free-fall time of close to 50 \%. 

To date, numerical simulations have generally chosen spherically symmetric or period boxes initial conditions of gravitationally bound clouds which
collapse and fragment to form stellar clusters \citep*{KleBurBat1998,BatBonBro2003,Bate2009a}. Cluster formation proceeds through hierarchical fragmentation and production of a
somewhat distributed  population which undergoes a hierachical merger process from small subclusters to one final
cluster containing most of the stars (\citealt*{BonBatVin2003, Bate2009a}; \citealt{Federrathetal2010}).
One simple possibility is that if star formation occurs in regions of molecular clouds
that are globally unbound, then there is no reason for the stars that form from the fragmenting population to fall together to
form the large stellar cluster. Recent  work evaluating the boundness of molecular clouds
show that their masses are typically five times smaller than that to be virialised, implying that 
much of the present day star formation is  occurring in unbound molecular clouds\citep{Heyeretal2009}. Here we demonstrate that the outcome of a distributed or clustered
population can depend on whether the  region is, or is not, globally bound.

Gravitationally unbound clouds have been explored in a series of studies to investigate how this
relates to the efficiency of star formation (\citealt{ClaBon2004, Clarketal2005}; \citealt*{ClaBonKle2008}).
Low star formation efficiencies are commonly taken to imply that star formation is slow and that 
molecular clouds are long-lived entities, supported by some internal mechanism and
lasting for several tens of dynamical times. In contrast, unbound clouds can also produce low star formation efficiencies
on dynamical timescales due to the fact that only a fraction of the cloud becomes gravitationally bound
due to the turbulence and undergoes gravitational collapse and star formation.

In this paper, we explore the importance of the local gravitational binding  in one cloud and show that a single
cloud can produce both a distributed and a clustered population, and a range of star formation efficiencies, depending on the
local gravitational binding.

\section{Calculations}

\begin{figure*}
%\vspace{-0.5truein}
\centerline{\psfig{figure=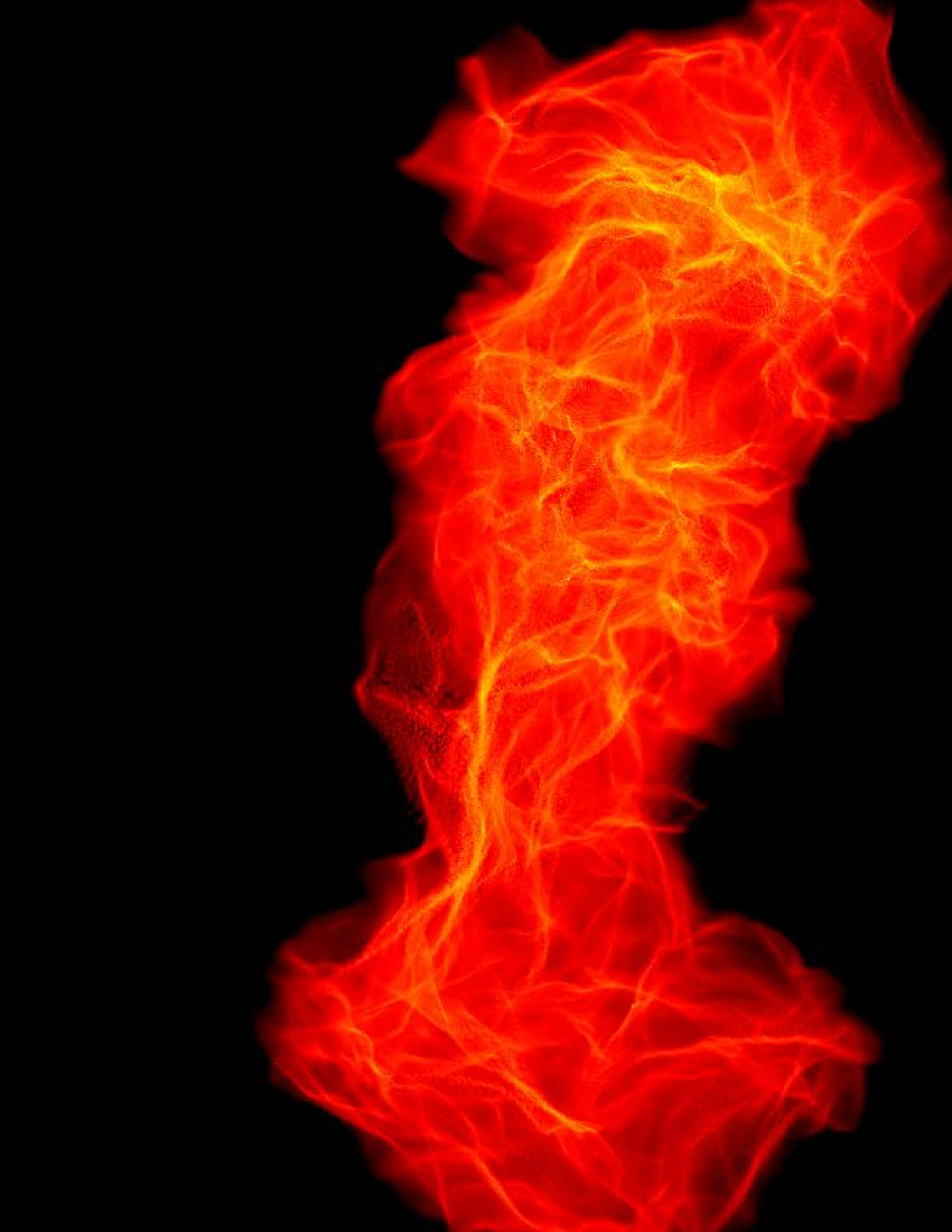,width=8.truecm,height=8.truecm}\hspace{0.1cm}\psfig{figure=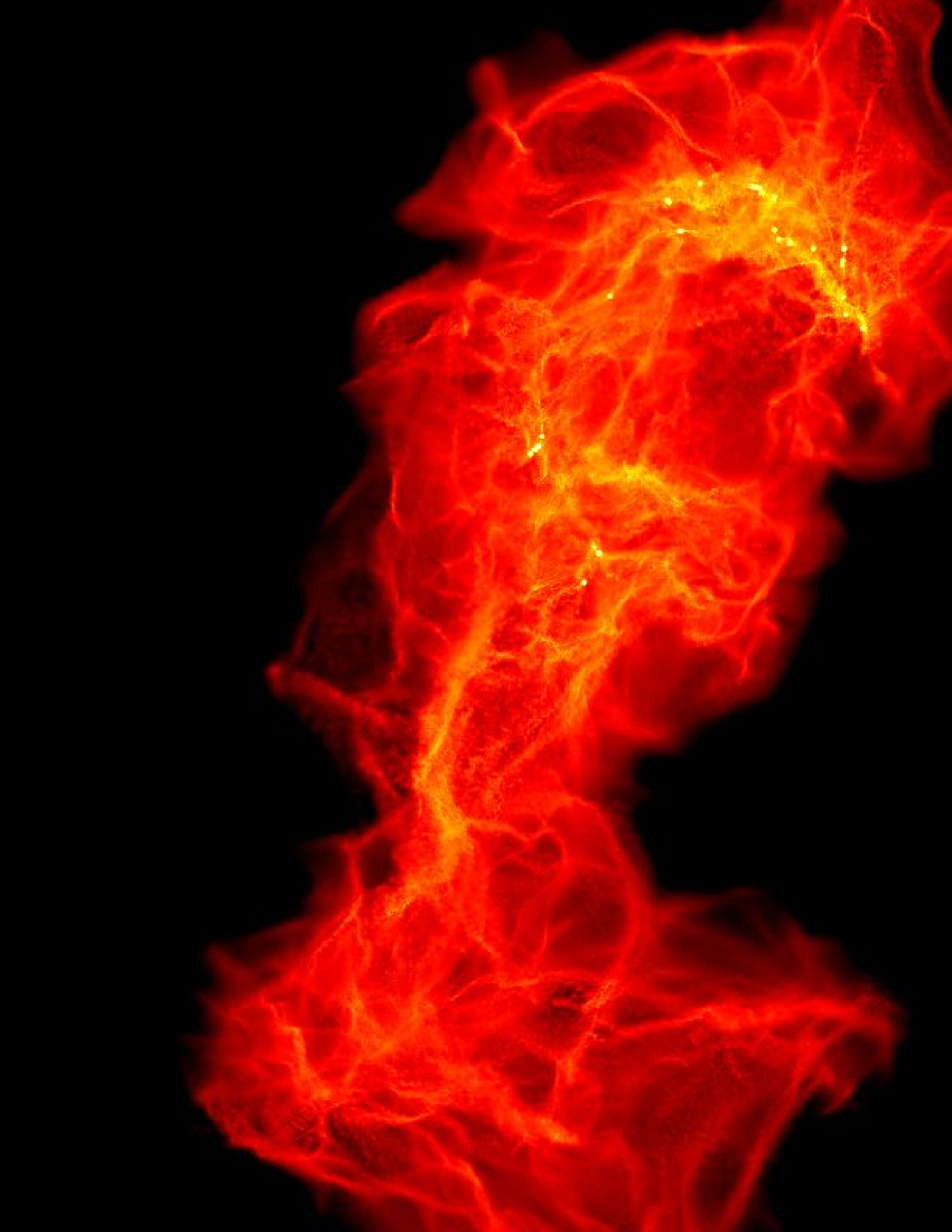,width=8.truecm,height=8.truecm}}\vspace{0.1cm}\centerline{\psfig{figure=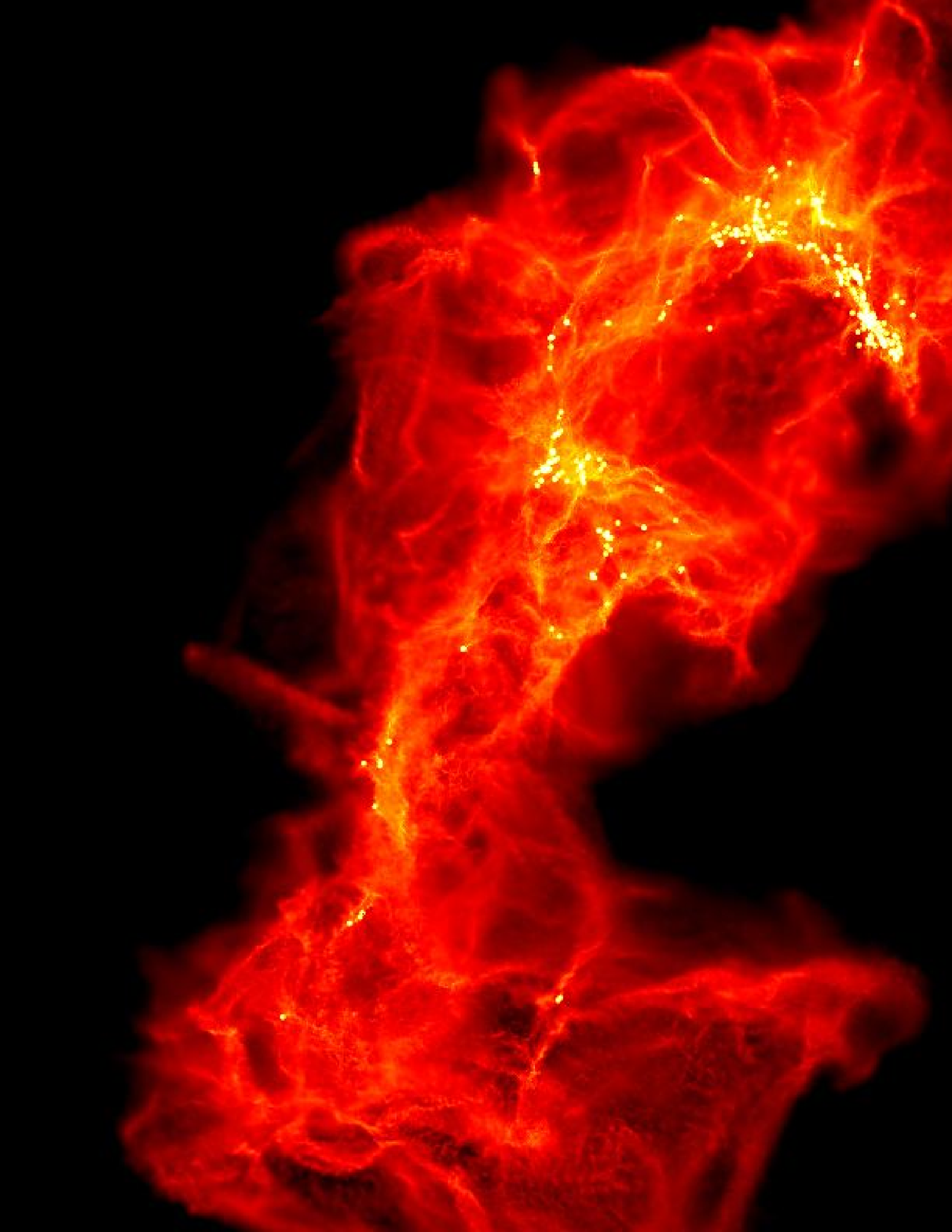,width=8.truecm,height=8.truecm}\hspace{0.1cm}\psfig{figure=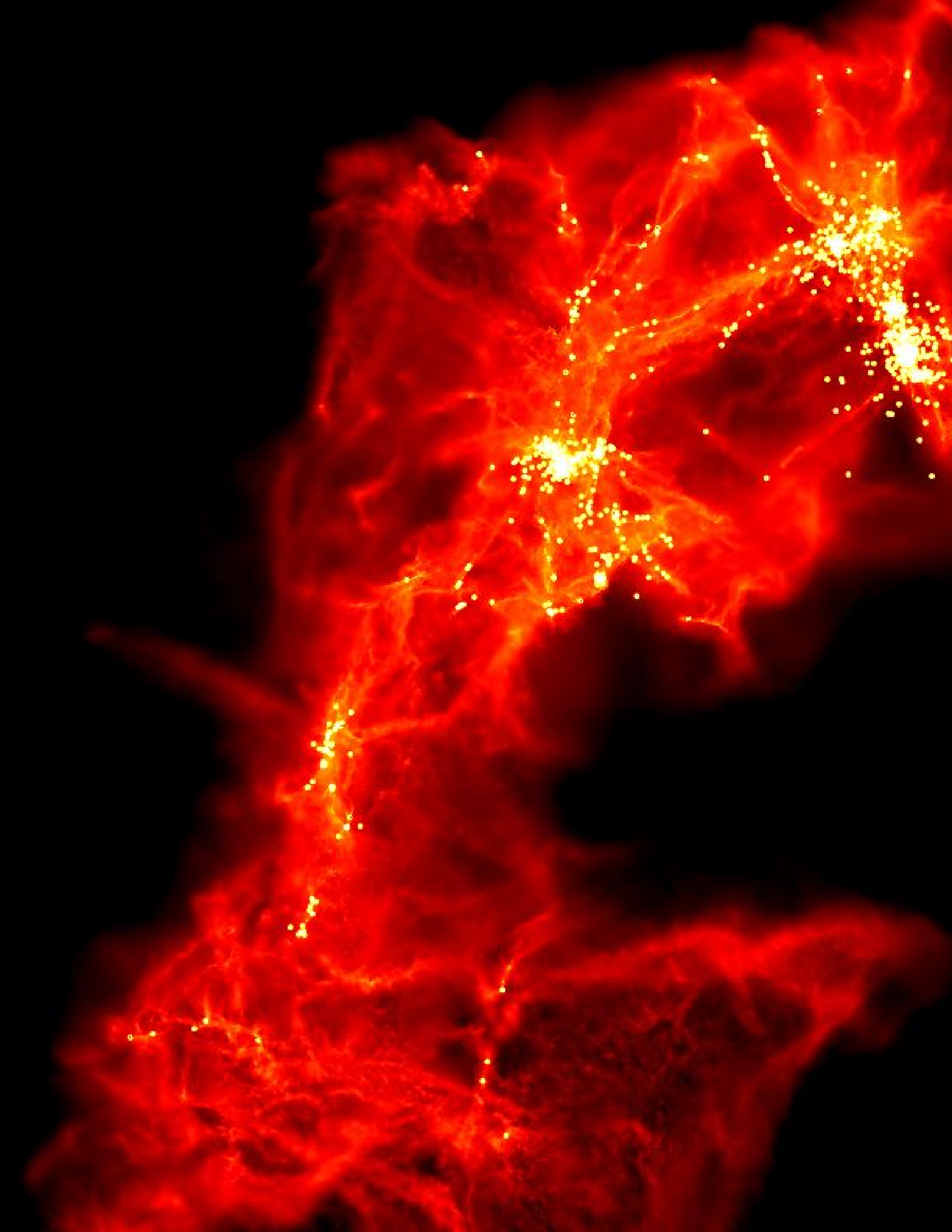,width=8.truecm,height=8.truecm}}
\caption{\label{OAevol}  The evolution of our model GMC as it evolves to form both a
distributed (bottom) and clustered (top) population of stars.  The cloud is initially
globally bound but a density gradient along the major axis
of the cloud makes the lower region unbound while the top region is gravitationally
bound. The cloud is shown at 0.365, 0.544, 0.727 0.961 $\tff$ ($\tff \approx 6.6 \times 10^5$) . Each panel shows the cloud in a 10x10pc region.
 The gas column densities are plotted on a logarithmic scale from 0.01 (black) to 100 (white) g~cm$^{-2}$.}
\end{figure*}

The results presented here are based on a large-scale Smoothed Particle Hydrodynamics (SPH)
simulation of a cylindrical $10^4$ \solmas\ molecular cloud 10 pc in length and 3 pc in cylindrical diameter. We have chosen an elongated cloud rather than the more standard spherical
cloud as most molecular clouds are non-sperhical and commonly elongated (e.g. Orion A).
Such a geometry can also produce additional structure due to gravitational focussing \citep{HarBur2007}.
This also allows for the physical properties to be varied along the cloud in a straightforward manner.
The cloud has a linear density gradient along its major axis with maximum/minimum values,
at each end of the cylinder,
$33$ percent high/lower than the average gas density of $ 1.35\times 10^{-20}$ g~cm$^{-3}$.
The gas has internal turbulence following a Larson-type $P(k) \sim k^{-4}$
power law throughout the cloud and is normalised such that the total kinetic energy balances the total gravitational energy in the cloud.
This corresponds to a full cloud (10 pc) 3-D velocity dispersion of order 4.5 km s$^{-1}$.
The density gradient applied then results in one end of the cloud being over bound (still super virial) while the other
end of the cloud is unbound. 

The cloud is populated with 15.5 million SPH particles on two levels,
providing high resolution in regions of interest. We initially performed a lower resolution
run with  5 million SPH particles producing an average mass resolution of $0.15 \solm$ \cite{BatBur1997}. Upon
completion of this low resolution simulation, we used three criteria to identify the regions that required higher resolution. This included the  particles which formed sinks, and those that were accreted onto sinks. It also included particles which attained sufficiently high density such that their local Jeans
mass was no longer resolved in the low-resolution run. All of these particles were identified and 
from the initial conditions of the low resolution run, they were split into 9 particles each to create the initial conditions for the high resolution simulations. This 
particle splitting was performed on the initial conditions to ensure that the physical quantities of mass, momentum,  
energy  and the energy spectrum were preserved. Note that the particle splitting does not introduce finer structure in the turbulent energy spectrum. This produced a mass resolution for the regions involved in star formation
of $0.0167 \solm$, sufficient to resolve the formation of higher-mass brown dwarfs, equivalent to a total
number of  $4.5 \times 10^7$ SPH particles. The equation of state (below) was specified in order to ensure that the Jeans mass in the higher resolution run did not descend below this mass resolution.

Particle splitting results in a marked increase in resolution without unmanageable computational costs \citep{KitWhi2002,KitWhi2007}. Note however some of the unsplit particles, which  in the low resolution run neither exceeded their Jeans mass limit 
nor became involved in the star formation, did get accreted by the additional stars in the high resolution run. This is to be expected as there are now additional locations of of star formation not present in
the low resolution run and these additional sinks will necessarily accrete unsplit particles.

The simulation follows a modified Larson-type equation of state \cite{Larson2005} comprised
of three barotropic equations of state 
\begin{equation}
P = k \rho^{\gamma}
\end{equation}
 where
\begin{equation}
\begin{array}{rlrl}
\gamma  &=  0.75  ; & \hfill &\rho \le \rho_1 \\
\gamma  &=  1.0  ; & \rho_1 < & \rho  \le \rho_2 \\
\gamma  &=  1.4  ; & \hfill \rho_2 < &\rho \le \rho_3 \\
\gamma  &=  1.0  ; & \hfill &\rho > \rho_3, \\
\end{array}
\end{equation}
and $\rho_1= 5.5 \times 10^{-19} {\rm g\ cm}^{-3} , \rho_2=5.5 \times 10^{-15} {\rm g\ cm}^{-3} , \rho_3=2 \times 10^{-13} {\rm g\ cm}^{-3}$.
 
The initial cooling part of the equation of state mimics the effects of line cooling and ensures that the
Jeans mass at the point of fragmentation is appropriate for characteristic stellar mass \citep{Jappsenetal2005, BonClaBat2006}.
The $\gamma=1.0$ approximates the effect of dust cooling while the $\gamma=1.4$ mimics the effects of an optically thick (to IR radiation) core, although its location at $\rho= 5.5 \times 10^{-15} {\rm g\ cm}^{-3}$, at lower densities than is typical, is in order to ensure that
the Jeans mass is always fully resolved and that a single self-gravitating fragment is turned into a sink particle. A higher critical density for this optically-thick phase where heating occurs would likely result in an increase in the numbers of low mass objects formed. The physical processes described would be unchanged. The final isothermal phase of the equation of state is simply in order to allow
sink-particle formation to occur,  which requires a subvirial collapsing fragment.
The initial conditions of the cloud contain 891 thermal Jeans masses ($M_{\rm Jeans} \approx  11 \solm$) such that if the cloud were
isothermal,
% and that the cloud was sufficiently bound that all the gas would eventually be involved in star formation,
 we would expect of order 900 fragments to form. 

Star formation in the cloud is modelled through the introduction of sink-particles \citep*{BatBonPri1995}.
Sink-particles formation is allowed once  the gas density of a collapsing fragment reaches $\rho\ge 6.8 \times 10^{-14}$ g\ cm$^{-3}$ although the equation of state ensures that this requires $\rho\ge 2. \times 10^{-13} {\rm g\ cm}^{-3}$. The neighbouring SPH
particles need be within a radius of $1. \times 10^{-3}$ pc and that fragment must be subvirial and collapsing.
Once created, the sinks accrete bound gas within $1. \times 10^{-3}$ pc and all gas that comes within $2. \times 10^{-4}$ pc.
The sinks have their mutual gravitational interactions smoothed to $2. \times 10^{-4}$ pc or 40 au. No interactions
including binary or disc disruptions can occur within this radius. 
 
 We assume a 100 \% efficiency of star formation within our sink particles. This likely overestimates the
 efficiency that would result were feedback from massive stars included. It is worth noting that our gas densities
 and core sizes are similar to the continuum surveys (Andre ..) that would require a 100 \% conversion in order
 to obtain a mapping from the core mass function to the stellar IMF. Furthermore, Previous simulations
 including ionisation and winds \citep{Daleetal2005, DalBon2008} do not find a large change in the
 resultant masses or mass spectra.
 
\section{Star formation and the developing IMF}

\begin{figure*}
\centerline{\psfig{figure=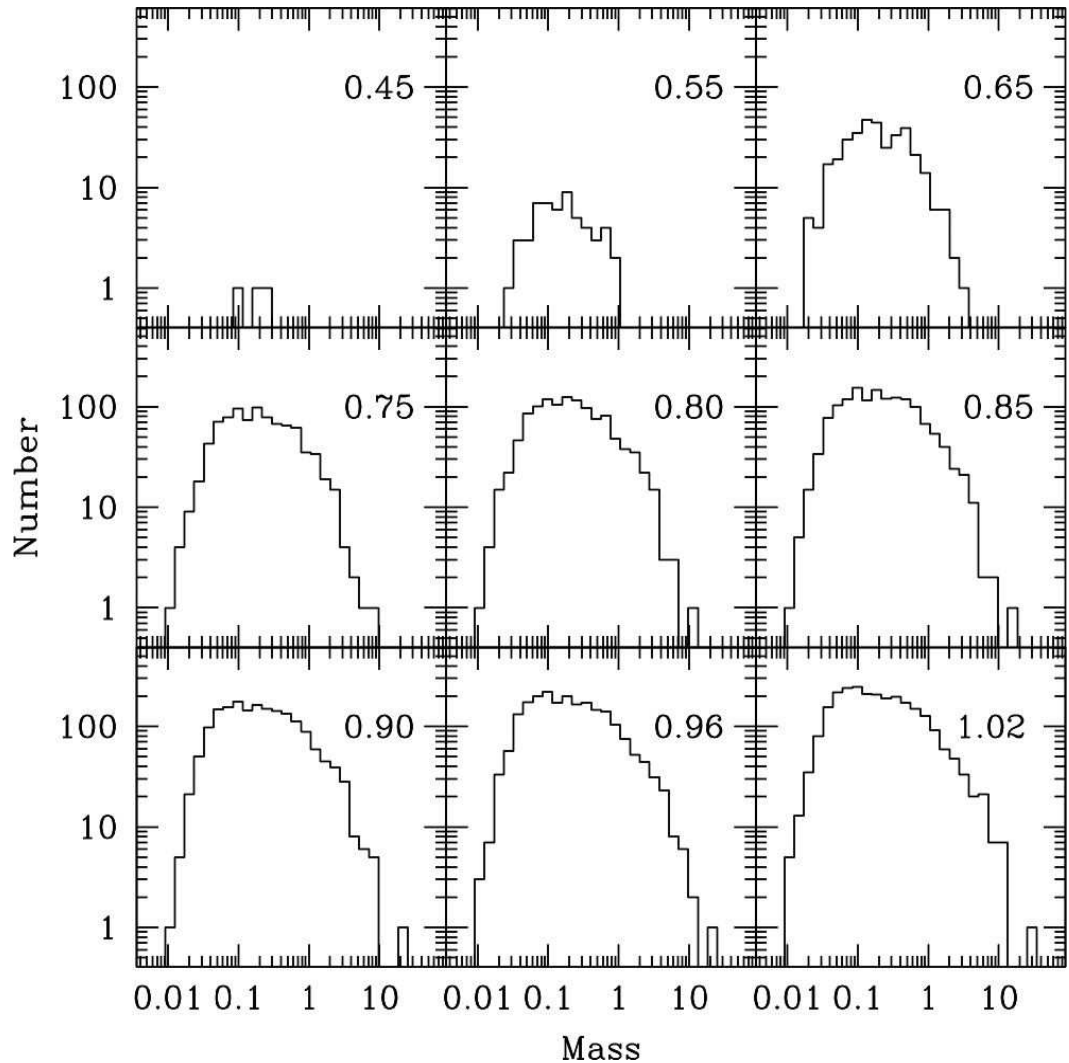,width=14.truecm,height=14.truecm}}
\caption{\label{imf9} The developing initial mass function of our star forming GC is plotted
at various times given in units of the clouds free-fall time, $\tff \approx 6.6 \times 10^5$ .  }
\end{figure*}

The simulation was followed for $1.02$ free-fall times or $\approx 6.6 \times 10^5$ years
and $\approx 3.9\times 10^5$ years after the first stars formed (see Fig.~\ref{OAevol}). During this time, 2542 stars
were formed with masses between $0.017$ and 30 \solmas.  The majority
of these stars form in the upper gravitationally bound part of the cloud while
some 7 per cent  form in the lower, gravitationally unbound regions. 

  Figure~\ref{imf9} shows the developing
initial mass function during the star formation process. The
stars form with masses comparable to the Jeans mass of the local gas. These initial masses
are initially of the order of several tenths of a Solar mass, while lower mass fragments form stars
later in the evolution due to the compression of gas to higher densities as it falls into
existing stellar clusters \citep*{BatBonBro2002a,BonClaBat2008}. Low and intermediate mass stars located in the centre
of forming clusters continue to accreted from the infalling gas and become high-mass stars \citep*[e.g.][]{  BonVinBat2004, SmiLonBon2009}. This produces a mass function that resembles the stellar IMF at all points during the
evolution with a continuous source of low mass stars forming with a decreasing subset of these accreting to ever higher masses. 
The high-mass end of the IMF is somewhat flatter than Salpeter \citep{Maschbergeretal2010}. This leaves room
for the additional physics of feedback from massive stars and the expected decrease in efficiency of massive star formation.
Note that by the massive stars attain their high-mass status through ongoing accretion over relatively long time-periods \citep{BonVinBat2004} such that their feedback could only affect the cloud after much of the star formation has occurred.

The cloud produces a variety of outcomes in terms of the distribution of stellar
masses, clustered and distributed modes of star formation as well as the
efficiency of the star formation process. These all depend largely on the 
initial conditions of the cloud and in particular to how gravitationally bound
the cloud is locally. The density gradient that is imposed along the major axis,
in conjunction with a constant specific kinetic energy of the gas, 
results in a local variation of the gravitational binding. Measured in terms of the
critical mass per unit length to be bound, this variation extends from 
$M/L=0.6$ (unbound) to $M/L=1.4$ (bound with the cloud overall
having a $M/L=1$.

\begin{figure*}
%\vspace{-0.5truein}
\centerline{\psfig{figure=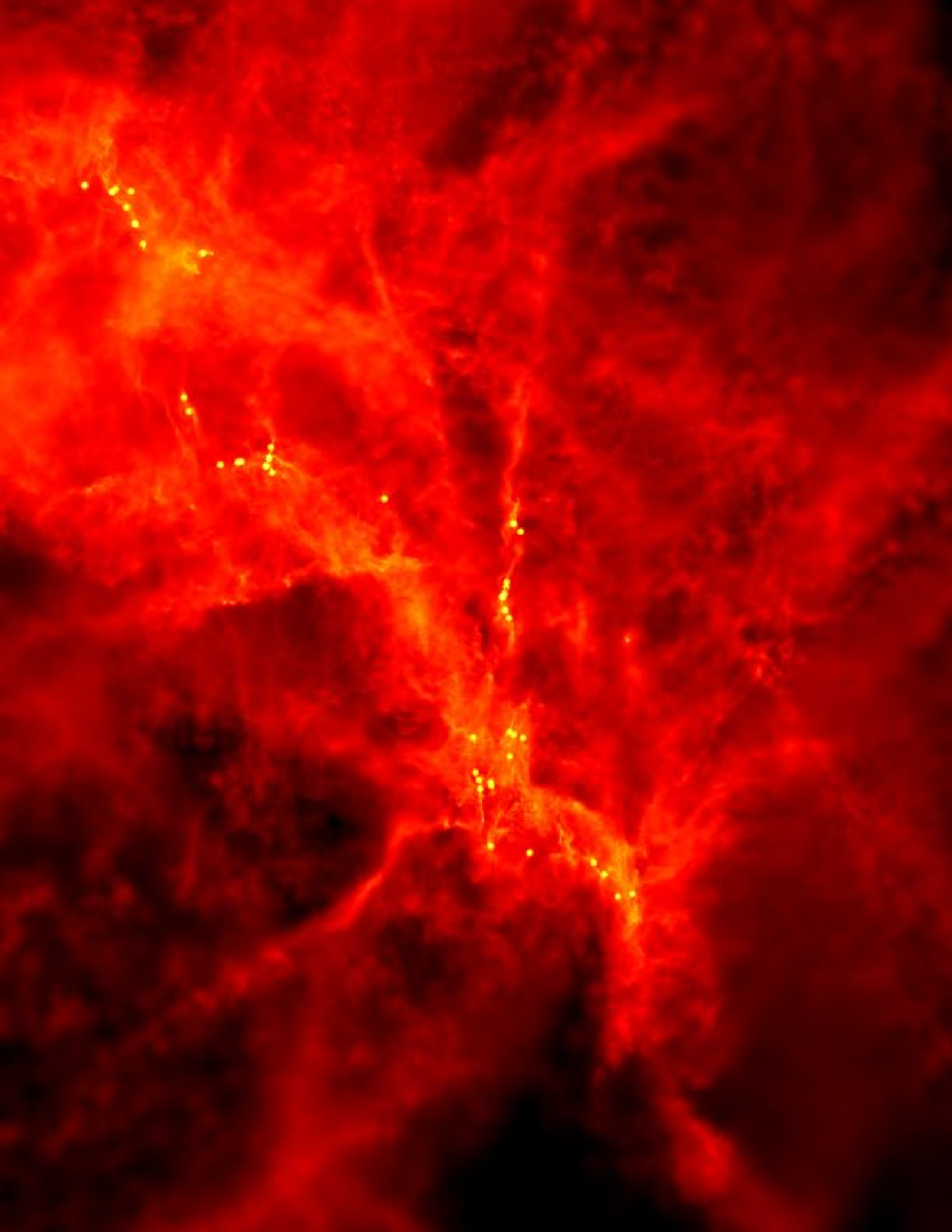,width=8.truecm,height=8.truecm}\hspace{0.1cm}\psfig{figure=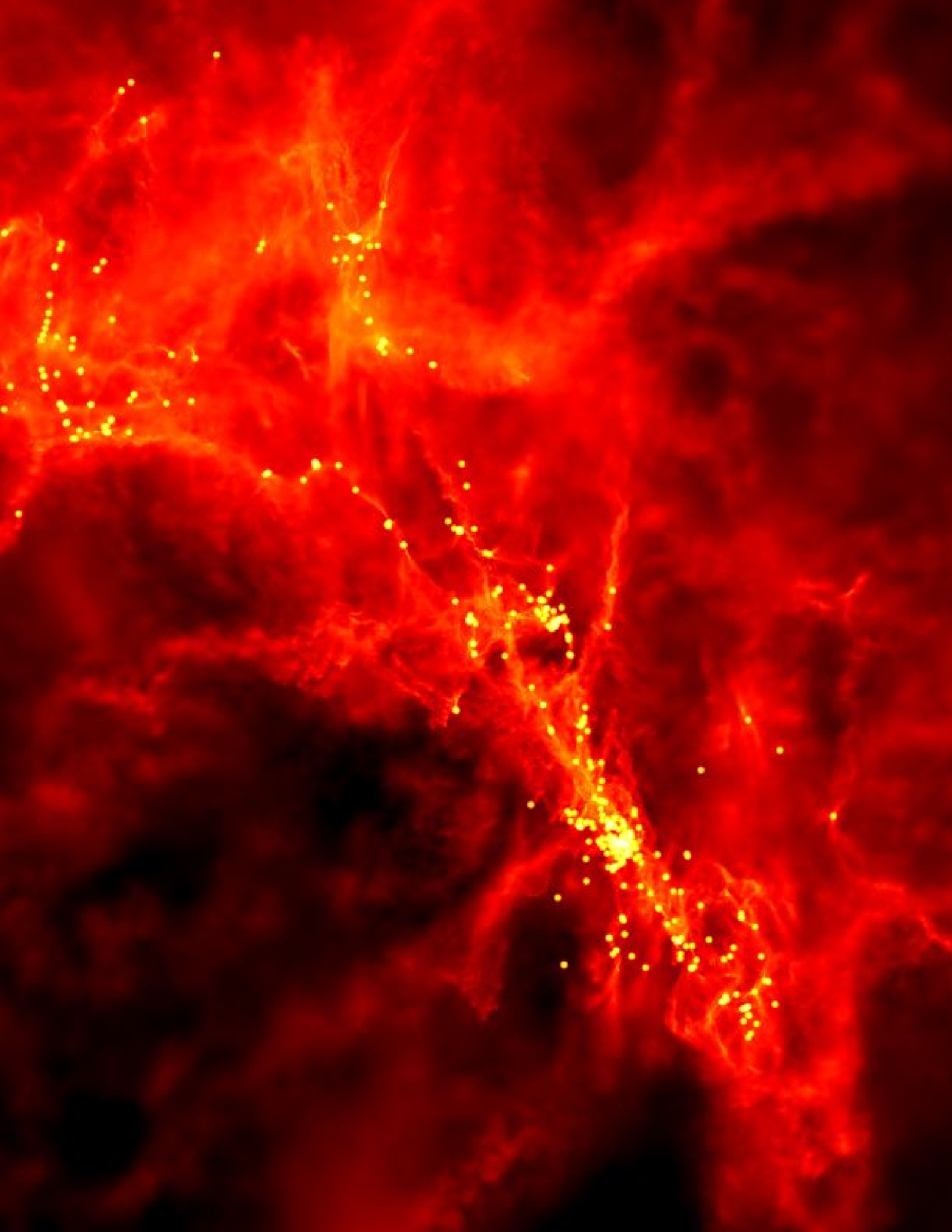,width=8.truecm,height=8.truecm}}
\vspace{0.1cm}\centerline{\psfig{figure=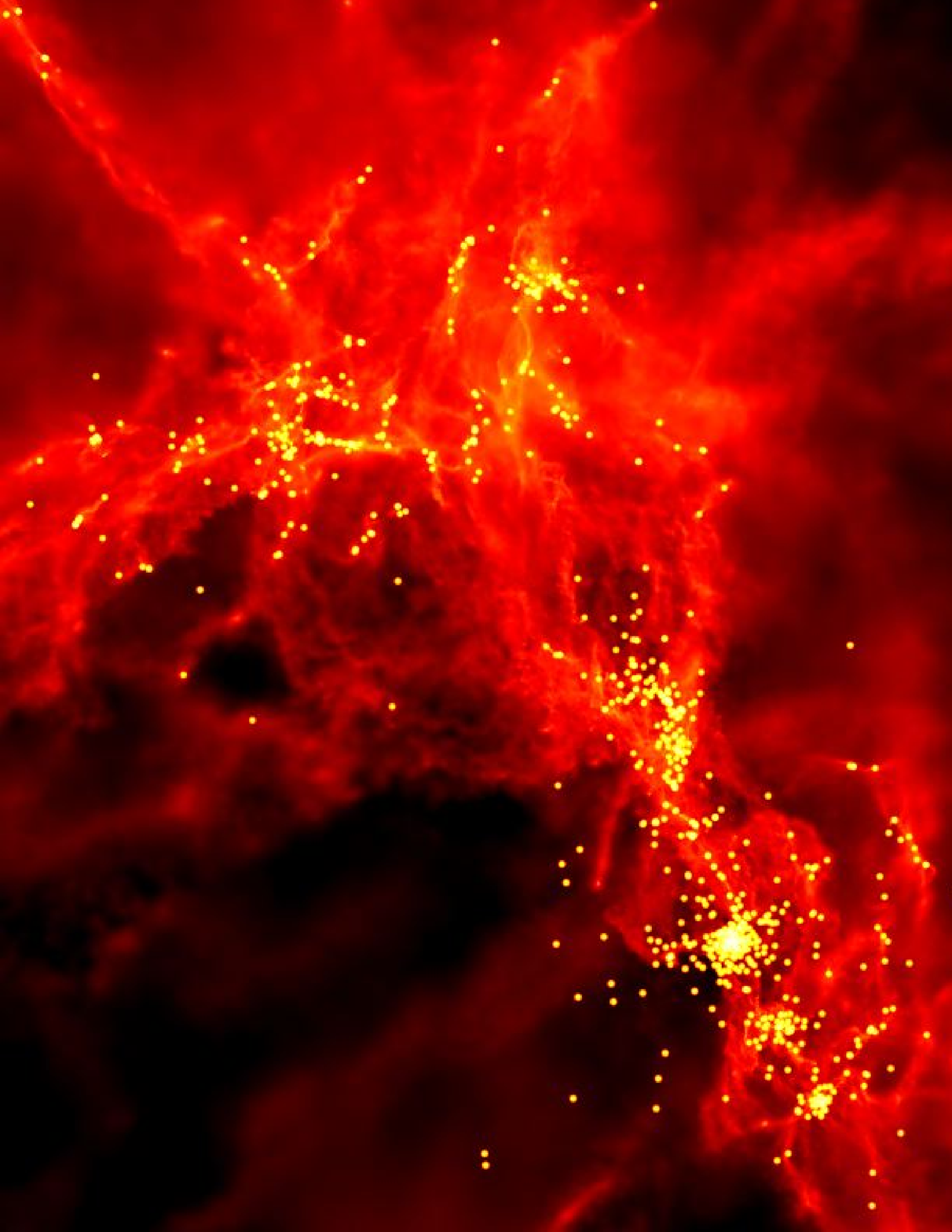,width=8.truecm,height=8.truecm}\hspace{0.1cm}\psfig{figure=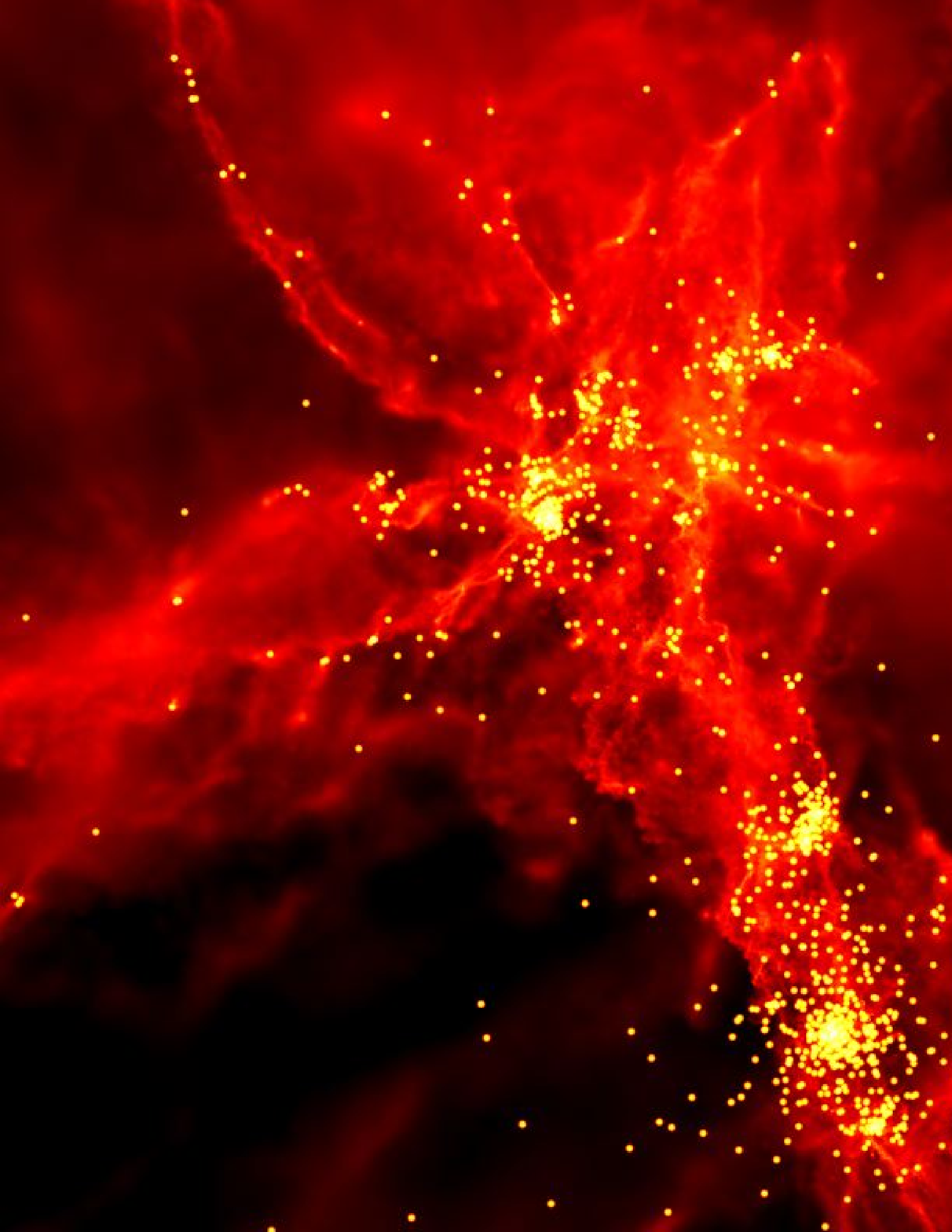,width=8.truecm,height=8.truecm}}
\caption{\label{OAclusform}  The formation of stellar clusters  occurs as infalling gas filaments fragment
to form small-N clusters. these clusters grow through the accretion of gas and stars and the merger of subclusters.
The cloud is shown at  0.58, 0.73, 0.87 and 1.02  $\tff$ ($\tff=\approx 6.6 \times 10^5$). Each panel shows the cloud in a 2.0x2.0pc region.
 The gas column densities are plotted on a logarithmic scale from 0.01 (black) to 100 (white) g\ cm$^{-2}$.}
\end{figure*}

\section{Clustering}

The evolution produced a number of high density clusters as well as a distributed population
of stars. The clusters form predominantly in the (upper)
bound regions of the cloud. The clusters form through the fragmentation of local overdense filamentary
structures that arise due to the turbulence, especially where such filaments intersect. Stars fall into local potential wells and form
small-N clusters which quickly grow by accreting other stars (and gas) that flow
along the filaments into the cluster potential. The merger of clusters also contributes to the growth 
of a stellar cluster. 
 Figure~\ref{OAclusform}  shows an example of this
process whereby accretion of gas and stars occurs along  filaments flowing into the cluster.

\begin{figure}
%\vspace{-0.5truein}
%\includegraphics[width=9.truecm]{densityvsMoLOrionAfn2.pdf}
\centerline{\psfig{figure=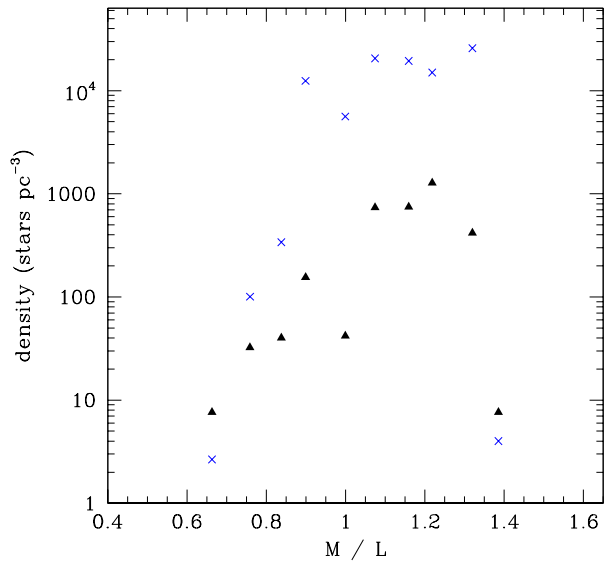,width=9.truecm,height=9.truecm,rheight=8.truecm}}
\caption{\label{clusvsMoL}  The stellar density is plotted as a function of the local gravitational binding measure in terms of the critical mass to length measure for a cylinder to be gravitationally bound. The higher (blue) points show the median density measured
from the ten nearest neighbours to each "star" while the lower value is measure over the local volume of the cloud. We see
that low stellar densities correspond to unbound regions of the cloud whereas high stellar densities result from bound regions.}
\end{figure}

An important result from this work is that the clustering depends strongly on the local gravitational binding of the gas prior to star formation. 
Figure~\ref{clusvsMoL} shows the resulting stellar densities as a function of how bound the cloud was
initially in terms of the critical mass per unit length, $M/L$, to be bound. Two measures of the stellar densities are plotted. The blue crosses show the median stellar density determined by the volume needed to contain the ten nearest stellar neighbours.  The black triangles show the density of stars contained in a fixed volume of size 0.5pc. The density determined by the first method is significantly higher as it typically is based on much smaller volumes. In both cases, the stellar density is low
in regions that were initially unbound and is much higher in the bound parts of the cloud with $M/L>1$.

This result is understandable in that clusters are (at least temporarally) bound objects and their formation requires that the pre-star formation gas is also bound. In locally unbound regions, it
is still possible to form small stellar clusters in regions where turbulent compression and shocks
result in a locally bound region. This process is more efficient in regions that are globally bound. Larger
scale regions containing many subsystems are bound even before any turbulent support is dissipated.
The smaller systems that form locally  can then hierarchically merge to form large stellar clusters 
\citep{BonBatVin2003, Bate2009a}. Residual gas in these bound regions then falls into the
gravitational potential of the cluster to be competitively accreted by the growing massive stars
located in the bottom of the potential well \citep{BonVinBat2004,BonBat2006}.
The massive stars are thus located in the  stellar clusters.

\begin{figure}
%\vspace{-0.5truein}
\centerline{\psfig{figure=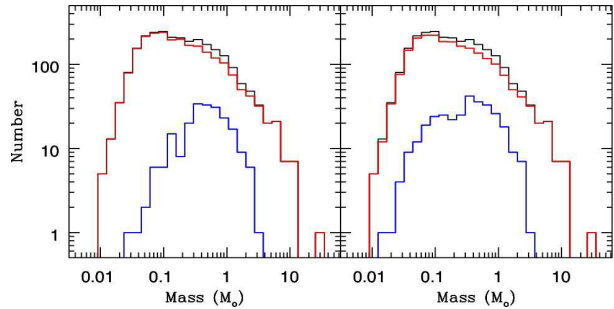,width=8.truecm,height=4.0truecm}}
\caption{\label{imfdenscut}  The initial mass functions at the end of the simulation are plotted for the total (top, black line)
population and for those stars formed in high-density regions (over 100 stars pc$^{-3}$, red line) and low density regions (lower, blue line). We see
that the low density regions do not form either high or low mass stars and thus result in an unusual IMF. The right
hand panel shows the same three distributions but where the stellar density is measured at the end of the simulation.
Several stars formed in high-density regions have been ejected from the clusters into the field. }
\end{figure}

\begin{figure}
%\vspace{-0.5truein}
\centerline{\psfig{figure=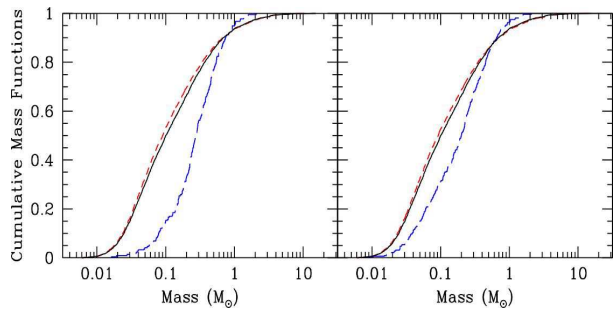,width=8.truecm,height=4.0truecm}}
\caption{\label{cimfdenscut}  The cumulative mass functions  are plotted for the total (solid black line)
population and for those stars formed in high-density (over 100 stars pc$^{-3}$, short-dashed red line) and low density (long-dashed blue line) regions.  The right
hand panel shows the same three distributions but where the stellar density is measured at the end of the simulation.
We see the low-density population is significantly different from that formed in the stellar clusters. This
difference remains even after stars are lost from the clusters into the low-density regions.}
\end{figure}

The majority of the stars and brown dwarfs formed are in high-density regions or have been ejected from stellar
clusters through interactions \citep{BatBonBro2002a,BatBonBro2003}. As noted in \citep{BonClaBat2008}, the brown dwarfs
predominantly form in stellar clusters due to the compression of the gas to high local densities
as it falls into the gravitational potential. The high-mass stars are also
predominantly formed in clusters \citep{BonVinBat2004, SmiLonBon2009}. This leads to a potentially observable
difference in the stellar IMFs of distributed and clustered star formation.
Figure~\ref{imfdenscut} shows the final IMF for the overall population and also
for distributed and clustered populations defined as those with a stellar density
lower or higher than 100 stars pc$^{-3}$. The right-hand panel of figure~\ref{imfdenscut}
shows the corresponding populations where they are separated by their maximum stellar
density during their evolution. The cumulative distributions (figure~\ref{cimfdenscut}) show
that the two distributions are statistically different and inconsistent (at the $1\times 10^{-13}$ level) with being drawn from the
same population. The distributed population has a significantly higher median stellar mass
and a pronounced lack of low-mass objects. This result helps explain the seemingly anomolous IMF
in Taurus which appears to have a lack of brown dwarfs and high-mass stars in a distributed population \citep{Luhman2004b}.

\section{Efficiency of star formation}

\begin{figure}
%\vspace{-0.5truein}
\centerline{\psfig{figure=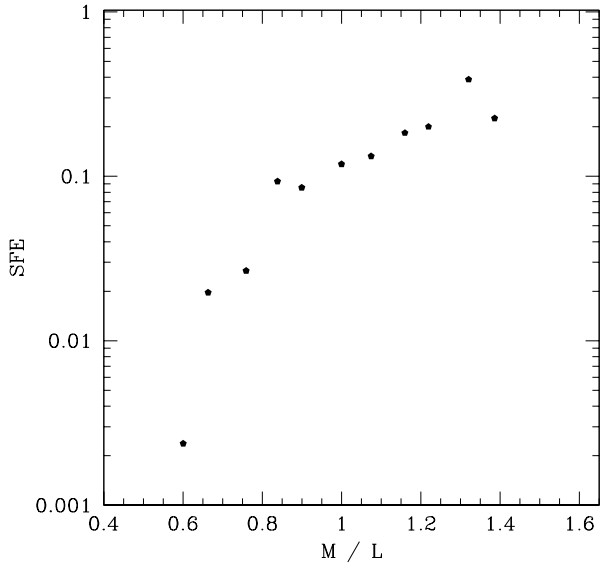,width=9.truecm,width=9.truecm,height=9.truecm,rheight=8.truecm}}
\caption{\label{sfevsmol} The star formation efficiency is plotted as a function of the critical mass to length ratio for a cylinder to be gravitationally bound. High star formation efficiencies result from very bound regions whereas very low
star formation efficiencies result in unbound regions. A mixture of such regions can thus result in the formation
of high SFE clustered regions and low SFE distributed populations.  }
\end{figure}

One of the central questions we wish to address in this paper is the relationship
between the nature and efficiency of the star formation process. Previous
studies \citep[i.e.][]{ClaBon2004, Clarketal2005, ClaBonKle2008} showed that unbound
clouds resulted in inefficient star formation. Furthermore, the efficiency reduces
dramatically the further the clouds are from being bound. Turbulent compression and shocks
results in some star formation  in these clouds but it is localised and much of the cloud escapes
without entering the star formation process. 

In the present study, we have one cloud that has regions which are bound and regions which
are unbound with a spatially varying $M/L$ from 0.6 to 1.4. This results in a range in 
local star formation efficiencies from 0.006 to 0.4. Figure~\ref{sfevsmol} plots the
local star formation efficicency as a function of the local binding of the cloud
in terms of the critical $M/L$ for the cloud to be globally bound. We  see that
after $1.02$ free-fall times or $\approx 6.6 \times 10^5$ years
(and $\approx 3.9\times 10^5$ years after the first stars formed) the local efficiency
of star formation is strongly dependent on local binding of the cloud. 
The bound regions have efficiencies varying from $10\%$ where the cloud is just bound to 
$20\%$ for a $M/L=1.2$ to a peak value of $40\%$ near where the cloud is maximally bound.
On the  unbound side, the efficiency quickly drops below $10\%$, reaching values
as low as a few \% for $M/L<0.8$. We can thus conclude that small changes in the local binding of the cloud
result in vastly different outcomes in terms of the star formation efficiency. 

It is worth noting that there is a strong correlation between the local efficiency of star formation and the
formation of stellar clusters. Regions  with relatively high efficiencies of $\ge 10$\% corresponds to regions
which are bound and thus form stellar clusters. In contrast the unbound regions form a  relatively distributed,
low stellar density population and does so at very low efficiencies. This is in agreement with observations where
clustered regions are found to have higher star forming efficiencies, whereas distributed regions such as Taurus
have low star formation efficiencies.  
These differences in the star formation efficiencies reported here
are not simply due to delays in star formation in the unbound parts of the cloud as most of the mass
is actually leaving the cloud and cannot partake in the star formation process \citep{ClaBonKle2008}.

\section{Observable determination of the star formation efficiency}

\begin{figure}
%\vspace{-0.5truein}
\centerline{\psfig{figure=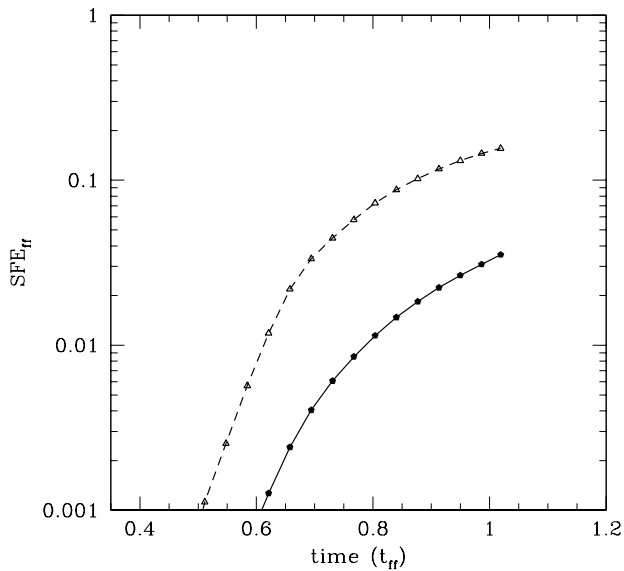,width=9.truecm,width=9.truecm,height=9.truecm,rheight=8.truecm}}
\caption{\label{sfevstime}  The  total (dashed) and time-averaged (solid) star formation efficiencies are plotted as a function of time.
While the final instantaneous SFE is of order 15 \%, an observer detecting a full population of such clouds would likely
sample all evolutionary stages and thus measure the much lower time-averaged star formation efficiency of $\approx 4$\%.}
\end{figure}

The evolution of the star formation efficiency is shown  in figure~\ref{sfevstime} from where the
first stars form at $t \approx 0.5 \tff$ to the end of the simulation at $t \approx 1.0 \tff$.
The overall star formation efficiency as measured at the end of the simulation is $\approx 15 $ \%.
this global value is an upper limit as no feedback effects are included \citep[e.g.][]{DalBon2008}. Magnetic fields could
also act to reduce this number further \citep{PriBat2008, PriBat2009}.  For example, if feedback
acted to destroy the cloud at $0.7 \tff$, then the final star formation efficiency would be of order $5$\%.

Such global star formation efficiencies are commonly invoked to discriminate between {\sl slow},  and {\sl fast}
star formation. Slow star formation invokes some supporting mechanism to prolong the lifetime of molecular
clouds to many tens of dynamical times \citep{KruTan2007} whereas fast star formation is expected to occur on timescales of several dynamical times \citep{Elmegreen2000}. Comparing such ideas to observations of large scale star formation rates
can be problematic as we do not know if the measured star formation efficiencies are final, intermediate or even pre-star formation values. Figure~\ref{sfevstime} also shows the time averaged star formation efficiency at any given time during the
cloud's evolution. This takes into account that when measuring global volume averaged values, we are just
as likely to observe any particular cloud at any point in its evolution. The time averaged star formation efficiency
is significantly lower than the instantaneous value throughout its evolution, due to the amount
of time the cloud spends in its pre-star formation stage. This neglects the time take for the cloud to form
which should be of order or greater than the dynamical time of the cloud. Nevertheless, the final time-averaged
star formation efficiency  is $\approx 3-4$\%. So, in a volume where a mixture of clouds of different evolutionary stages
are present, an observer would estimate  star formation efficiency of several percent and  thus conclude that 
this was {\sl slow} star formation, when in reality star formation is locally proceeding on a dynamical timescale.
All of the above neglects the effects of magnetic fields and feedback which both act to reduce the star formation
rates and efficiencies.  While magnetic fields do slow down star formation \citep{PriBat2008, PriBat2009}, feedback
impedes or locally stops star formation without appreciably changing its dynamical nature \citep{Daleetal2005, DalBon2008}

\section{Discussion: The Evolution of Giant Molecular Clouds}

We have seen from the above that low star formation efficiencies are plausibly due to the GMCs not being completely bound by their self gravity. Even in the absence of effects such as magnetic fields and feedback, clouds that are globally, or at least in large part  unbound result in low star formation efficiencies. Star formation requires that the clouds are nearly bound or at least have 
significant regions
which are close to being bound, with typically gravitational and kinetic energies within a factor of a few of each other (\citealt*{ClaBonKle2008}). From this starting point, we can try to construct an evolutionary scenario for GMCs in which
self-gravity only plays a role in the actual star formation process \citep{Pringleetal2001, Bonnelletal2006, Dobbsetal2006}.
We will neglect the effects of the magnetic field but to first order its effects would be to increase the internal energy
of the clouds such that they are further from being gravitationally bound.

The formation of giant molecular clouds is uncertain with various theories from gravitational instabilities to
cloud coagulation and spiral shocks \citep{MckeeOstriker2007}. In reality all of these processes may play a significant role in GMC
formation but for our purposes, we will assume that GMC formation occurs primarily due to spiral shocks or
cloud coagulations \citep{Dobbsetal2006}. 
In such cases, self-gravity need not play an important role in the formation process and clouds
can be formed without being gravitationally bound. 

Cloud formation from external collisions or compression implies an increasing contribution to the cloud's gravitational potential.
During formation, the cloud evolves from a state where self-gravity is unimportant to one where self-gravity has a significant effect
on the cloud's dynamics. Once the gravitational energy is within a factor of a few of the kinetic energy, then star formation
proceeds in local regions that become bound due to the cloud's internal dynamics (\citealt*{ClaBonKle2008}).
This local star formation will occur as parts of the cloud are still being assembled as the local timescale is much
shorter than the overall dynamical time for the cloud or its precursor. 

Star formation would then proceed until either the local gas reservoir is depleted, or until the gas reservoir
is removed by the effects of feedback from young stars. The tidal shear from leaving the spiral arm potential
could also limit the lifetime of the clouds \citep{Dobbsetal2006}. 
The majority of the cloud need never become gravitationally bound before the cloud is dispersed resulting in
inefficient star formation process that is still occurring on a fast dynamical timescale.

\section{Conclusions}

Star formation in realistic GMCs will proceed from a variety of physical conditions, spanning regions
that are gravitationally bound to parts or whole clouds which are gravitationally unbound. 
Star formation will occur as long as the local conditions are close to being gravitationally bound
but the properties of the young stellar population can depend strongly on these conditions. Regions
that are bound produce bound stellar clusters and a stellar population that follows   the full initial mass function
from brown dwarfs to high-mass stars. Regions that are unbound are likely to produce a somewhat skewed
IMF biased towards the local Jeans mass in the gas and with significant lack of lower-mass stars, such
as is seen in Taurus \citep{Luhman2004b}. 

The star formation efficiency is also a product of the local physical conditions with bound regions resulting in a relatively
high star formation efficiency of order 10 \% or more per free-fall time. Regions that are unbound can have
drastically reduced efficiencies of order 1 \% or less per free-fall time. Thus clustered star formation should
occur in regions of higher local star formation efficiencies that more distributed populations. 

Estimates of low star formation effiiciencies are equally consistent with fast dynamical star formation as
slow quasistatic star formation provided that one relaxes the condition that GMCs are globally bound long-lived
entities. Including the pre-star formation timeperiods where clouds are being assembled, global estimates
of depletion timescales or star formation rates per free-fall time will appear to be low even while
local regions are undergoing fast star formation at high efficiencies.

Finally, realistic GMCs are likely to be constructed from a mix of physical conditions  such that 
a fraction of the cloud is bound producing stellar clusters at high efficiencies whereas the
majority of the cloud is unbound producing a more distributed population at low
star formation efficiencies before the cloud is unbound by feedback or alternative process.
Such a scenario is consistent with a model where GMCs are not formed due to their self-gravity
but rather to an external process such as spiral shocks  \citep{Dobbsetal2006, Dobbs2008}. 

\section*{Acknowledgments}
We acknowledge the contribution of the  U.K. Astrophysical Fluids Facility (UKAFF) and SUPA
for providing the computational facilities for the simulations reported here.  IAB thanks
the ETCC committee of STFC for providing the rail journeys on which this paper was written.
PCC. acknowledges support by the Deutsche Forschungsgemeinschaft (DFG) under grant KL 1358/5 and via the Sonderforschungsbereich (SFB) SFB 439, Galaxien im fr\"uhen Universum. 
MRB is grateful for the support of a Philip Leverhulme Prize and a EURYI Award.
This work, conducted as part of the award  ``The formation of stars
and planets: Radiation hydrodynamical and magnetohydrodynamical simulationsÕ' made under the European Heads of Research Councils and European Science Foundation EURYI (European Young
Investigator) Awards scheme, was supported by funds from the Participating Organizations of EURYI and the EC Sixth Framework
Programme.
We would like to thank Chris Rudge and Richard West at the UK Astrophysical Fluid Facility (UKAFF) for their tireless assistance and enthusiasm during the completion of this work.
%MRB is grateful for the support of a Philip Leverhulme Prize.
%and some other excessive european subsidy to ensure he does not toil fruitlessly in the classroom. 

%\bibliography{/Users/mbate/Documents/home2/Tex/Papers/Bonnelletal2010/mbate}
\bibliography{iab}

\begin{thebibliography}{}

\bibitem[\protect\citeauthoryear{{Allen}, {Megeath}, {Gutermuth}, {Myers},
  {Wolk}, {Adams}, {Muzerolle}, {Young} \& {Pipher}}{{Allen}
  et~al.}{2007}]{AllenMegeathetal2007}
{Allen} L.,  {Megeath} S.~T.,  {Gutermuth} R.,  {Myers} P.~C.,  {Wolk} S.,
  {Adams} F.~C.,  {Muzerolle} J.,  {Young} E.,    {Pipher} J.~L.,  2007,
  Protostars and Planets V, pp 361--376

\bibitem[\protect\citeauthoryear{{Bate}}{{Bate}}{2009}]{Bate2009a}
{Bate} M.~R.,  2009, \mnras, 392, 590

\bibitem[\protect\citeauthoryear{{Bate}, {Bonnell} \& {Bromm}}{{Bate}
  et~al.}{2002}]{BatBonBro2002a}
{Bate} M.~R.,  {Bonnell} I.~A.,    {Bromm} V.,  2002, MNRAS, 332, L65

\bibitem[\protect\citeauthoryear{{Bate}, {Bonnell} \& {Bromm}}{{Bate}
  et~al.}{2003}]{BatBonBro2003}
{Bate} M.~R.,  {Bonnell} I.~A.,    {Bromm} V.,  2003, MNRAS, 339, 577

\bibitem[\protect\citeauthoryear{{Bate}, {Bonnell} \& {Price}}{{Bate}
  et~al.}{1995}]{BatBonPri1995}
{Bate} M.~R.,  {Bonnell} I.~A.,    {Price} N.~M.,  1995, MNRAS, 277, 362

\bibitem[\protect\citeauthoryear{{Bate} \& {Burkert}}{{Bate} \&
  {Burkert}}{1997}]{BatBur1997}
{Bate} M.~R.,  {Burkert} A.,  1997, \mnras, 288, 1060

\bibitem[\protect\citeauthoryear{{Bonnell} \& {Bate}}{{Bonnell} \&
  {Bate}}{2006}]{BonBat2006}
{Bonnell} I.~A.,  {Bate} M.~R.,  2006, \mnras, 370, 488

\bibitem[\protect\citeauthoryear{{Bonnell}, {Bate} \& {Vine}}{{Bonnell}
  et~al.}{2003}]{BonBatVin2003}
{Bonnell} I.~A.,  {Bate} M.~R.,    {Vine} S.~G.,  2003, MNRAS, 343, 413

\bibitem[\protect\citeauthoryear{{Bonnell}, {Clark} \& {Bate}}{{Bonnell}
  et~al.}{2008}]{BonClaBat2008}
{Bonnell} I.~A.,  {Clark} P.,    {Bate} M.~R.,  2008, \mnras, 389, 1556

\bibitem[\protect\citeauthoryear{{Bonnell}, {Clarke} \& {Bate}}{{Bonnell}
  et~al.}{2006}]{BonClaBat2006}
{Bonnell} I.~A.,  {Clarke} C.~J.,    {Bate} M.~R.,  2006, \mnras, 368, 1296

\bibitem[\protect\citeauthoryear{{Bonnell}, {Dobbs}, {Robitaille} \&
  {Pringle}}{{Bonnell} et~al.}{2006}]{Bonnelletal2006}
{Bonnell} I.~A.,  {Dobbs} C.~L.,  {Robitaille} T.~P.,    {Pringle} J.~E.,
  2006, MNRAS, 365, 37

\bibitem[\protect\citeauthoryear{{Bonnell}, {Vine} \& {Bate}}{{Bonnell}
  et~al.}{2004}]{BonVinBat2004}
{Bonnell} I.~A.,  {Vine} S.~G.,    {Bate} M.~R.,  2004, \mnras, 349, 735

\bibitem[\protect\citeauthoryear{{Clark} \& {Bonnell}}{{Clark} \&
  {Bonnell}}{2004}]{ClaBon2004}
{Clark} P.~C.,  {Bonnell} I.~A.,  2004, \mnras, 347, L36

\bibitem[\protect\citeauthoryear{{Clark}, {Bonnell} \& {Klessen}}{{Clark}
  et~al.}{2008}]{ClaBonKle2008}
{Clark} P.~C.,  {Bonnell} I.~A.,    {Klessen} R.~S.,  2008, \mnras, 386, 3

\bibitem[\protect\citeauthoryear{{Clark}, {Bonnell}, {Zinnecker} \&
  {Bate}}{{Clark} et~al.}{2005}]{Clarketal2005}
{Clark} P.~C.,  {Bonnell} I.~A.,  {Zinnecker} H.,    {Bate} M.~R.,  2005,
  \mnras, 359, 809

\bibitem[\protect\citeauthoryear{{Clarke}, {Bonnell} \& {Hillenbrand}}{{Clarke}
  et~al.}{2000}]{ClaBonHil2000}
{Clarke} C.~J.,  {Bonnell} I.~A.,    {Hillenbrand} L.~A.,  2000, Protostars and
  Planets IV, pp 151--+

\bibitem[\protect\citeauthoryear{{Dale} \& {Bonnell}}{{Dale} \&
  {Bonnell}}{2008}]{DalBon2008}
{Dale} J.~E.,  {Bonnell} I.~A.,  2008, \mnras, 391, 2

\bibitem[\protect\citeauthoryear{{Dale}, {Bonnell}, {Clarke} \& {Bate}}{{Dale}
  et~al.}{2005}]{Daleetal2005}
{Dale} J.~E.,  {Bonnell} I.~A.,  {Clarke} C.~J.,    {Bate} M.~R.,  2005,
  \mnras, 358, 291

\bibitem[\protect\citeauthoryear{{Dobbs}}{{Dobbs}}{2008}]{Dobbs2008}
{Dobbs} C.~L.,  2008, \mnras, 391, 844

\bibitem[\protect\citeauthoryear{{Dobbs}, {Bonnell} \& {Pringle}}{{Dobbs}
  et~al.}{2006}]{Dobbsetal2006}
{Dobbs} C.~L.,  {Bonnell} I.~A.,    {Pringle} J.~E.,  2006, \mnras, 371, 1663

\bibitem[\protect\citeauthoryear{{Elmegreen}}{{Elmegreen}}{2000}]{Elmegreen200%
0}
{Elmegreen} B.~G.,  2000, \mnras, 311, L5

\bibitem[\protect\citeauthoryear{{Evans}, {Dunham}, {J{\o}rgensen}, {Enoch},
  {Mer{\'{\i}}n}, {van Dishoeck}, {Alcal{\'a}} \& {Myers}}{{Evans}
  et~al.}{2009}]{Evansetal2009}
{Evans} N.~J.,  {Dunham} M.~M.,  {J{\o}rgensen} J.~K.,  {Enoch} M.~L.,
  {Mer{\'{\i}}n} B.,  {van Dishoeck} E.~F.,  {Alcal{\'a}} J.~M.,    {Myers}
  P.~C.,  2009, \apjs, 181, 321

\bibitem[\protect\citeauthoryear{{Federrath}, {Banerjee}, {Clark} \&
  {Klessen}}{{Federrath} et~al.}{2010}]{Federrathetal2010}
{Federrath} C.,  {Banerjee} R.,  {Clark} P.~C.,    {Klessen} R.~S.,  2010,
  \apj, 713, 269

\bibitem[\protect\citeauthoryear{{Gutermuth}, {Megeath}, {Myers}, {Allen},
  {Pipher} \& {Fazio}}{{Gutermuth} et~al.}{2009}]{Gutermuthetal2009}
{Gutermuth} R.~A.,  {Megeath} S.~T.,  {Myers} P.~C.,  {Allen} L.~E.,  {Pipher}
  J.~L.,    {Fazio} G.~G.,  2009, \apjs, 184, 18

\bibitem[\protect\citeauthoryear{{Gutermuth}, {Myers}, {Megeath}, {Allen},
  {Pipher}, {Muzerolle}, {Porras}, {Winston} \& {Fazio}}{{Gutermuth}
  et~al.}{2008}]{Gutermuthetal2008}
{Gutermuth} R.~A.,  {Myers} P.~C.,  {Megeath} S.~T.,  {Allen} L.~E.,  {Pipher}
  J.~L.,  {Muzerolle} J.,  {Porras} A.,  {Winston} E.,    {Fazio} G.,  2008,
  \apj, 674, 336

\bibitem[\protect\citeauthoryear{{Hartmann} \& {Burkert}}{{Hartmann} \&
  {Burkert}}{2007}]{HarBur2007}
{Hartmann} L.,  {Burkert} A.,  2007, \apj, 654, 988

\bibitem[\protect\citeauthoryear{{Hillenbrand} \& {Hartmann}}{{Hillenbrand} \&
  {Hartmann}}{1998}]{HilHar1998}
{Hillenbrand} L.~A.,  {Hartmann} L.~W.,  1998, ApJ, 492, 540

\bibitem[\protect\citeauthoryear{{Jappsen}, {Klessen}, {Larson}, {Li} \& {Mac
  Low}}{{Jappsen} et~al.}{2005}]{Jappsenetal2005}
{Jappsen} A.-K.,  {Klessen} R.~S.,  {Larson} R.~B.,  {Li} Y.,    {Mac Low}
  M.-M.,  2005, \aap, 435, 611

\bibitem[\protect\citeauthoryear{{Kitsionas} \& {Whitworth}}{{Kitsionas} \&
  {Whitworth}}{2002}]{KitWhi2002}
{Kitsionas} S.,  {Whitworth} A.~P.,  2002, \mnras, 330, 129

\bibitem[\protect\citeauthoryear{{Kitsionas} \& {Whitworth}}{{Kitsionas} \&
  {Whitworth}}{2007}]{KitWhi2007}
{Kitsionas} S.,  {Whitworth} A.~P.,  2007, \mnras, 378, 507

\bibitem[\protect\citeauthoryear{{Klessen}, {Burkert} \& {Bate}}{{Klessen}
  et~al.}{1998}]{KleBurBat1998}
{Klessen} R.~S.,  {Burkert} A.,    {Bate} M.~R.,  1998, ApJ, 501, L205+

\bibitem[\protect\citeauthoryear{{Krumholz} \& {Tan}}{{Krumholz} \&
  {Tan}}{2007}]{KruTan2007}
{Krumholz} M.~R.,  {Tan} J.~C.,  2007, \apj, 654, 304

\bibitem[\protect\citeauthoryear{{Lada} \& {Lada}}{{Lada} \&
  {Lada}}{2003}]{LadLad2003}
{Lada} C.~J.,  {Lada} E.~A.,  2003, \araa, 41, 57

\bibitem[\protect\citeauthoryear{{Lada}, {Depoy}, {Evans} II \&
  {Gatley}}{{Lada} et~al.}{1991}]{Ladaetal1991}
{Lada} E.~A.,  {Depoy} D.~L.,  {Evans} II N.~J.,    {Gatley} I.,  1991, \apj,
  371, 171

\bibitem[\protect\citeauthoryear{{Larson}}{{Larson}}{2005}]{Larson2005}
{Larson} R.~B.,  2005, \mnras, 359, 211

\bibitem[\protect\citeauthoryear{{Luhman}}{{Luhman}}{2004}]{Luhman2004b}
{Luhman} K.~L.,  2004, \apj, 617, 1216

\bibitem[\protect\citeauthoryear{{Maschberger}, {Clarke}, {Bonnell} \&
  {Kroupa}}{{Maschberger} et~al.}{2010}]{Maschbergeretal2010}
{Maschberger} T.,  {Clarke} C.~J.,  {Bonnell} I.~A.,    {Kroupa} P.,  2010,
  \mnras, 404, 1061

\bibitem[\protect\citeauthoryear{{McKee} \& {Ostriker}}{{McKee} \&
  {Ostriker}}{2007}]{MckeeOstriker2007}
{McKee} C.~F.,  {Ostriker} E.~C.,  2007, \araa, 45, 565

\bibitem[\protect\citeauthoryear{{Price} \& {Bate}}{{Price} \&
  {Bate}}{2008}]{PriBat2008}
{Price} D.~J.,  {Bate} M.~R.,  2008, \mnras, 385, 1820

\bibitem[\protect\citeauthoryear{{Price} \& {Bate}}{{Price} \&
  {Bate}}{2009}]{PriBat2009}
{Price} D.~J.,  {Bate} M.~R.,  2009, \mnras ~submitted, 0

\bibitem[\protect\citeauthoryear{{Pringle}, {Allen} \& {Lubow}}{{Pringle}
  et~al.}{2001}]{Pringleetal2001}
{Pringle} J.~E.,  {Allen} R.~J.,    {Lubow} S.~H.,  2001, \mnras, 327, 663

\bibitem[\protect\citeauthoryear{{Smith}, {Longmore} \& {Bonnell}}{{Smith}
  et~al.}{2009}]{SmiLonBon2009}
{Smith} R.~J.,  {Longmore} S.,    {Bonnell} I.,  2009, \mnras, 400, 1775

\end{thebibliography}

\bsp

\label{lastpage}

\end{document}